# RNA Detection in air by means of Cosmic Rays interactions

Adlish J. I[a]., Mainardi E[a]., Neuhold P[a]., Surrente R[a]., Tagliapietra L. J[a].

*BabuHawaii Foundation, a US Public Entity on Scientific Research*[a]
*5940 S Rainbow Blvd Ste 400, Las Vegas, Nevada 89118 USA*[a]

Correspondence to: Piero Neuhold piero@babuhawaiifoundation.org

**Abstract**

The study research presented hereafter shows a new methodology to reveal traces of Viral particles thanks to their own chemical structure such as Prosperous, an element making up part of the structure of RNA, a type of nucleic acid, such as in a virus, in an open space or a closed ambient ( typically an airport hall)  detecting sub-atomic particles interactions with the air due to Cosmic Rays , an Outer Space free source .This is constructed with the use of adequate detectors looking in particular at the peculiar interactions of muons, cosmic rays relativistic particle segment, with the no living matter present as aerosol in the air.

1. **Introduction**

Cosmic rays [19,20], coming from outer space and travelling almost at the speed of light, produce a huge cascade of different relativistic particles interacting with air molecules at high altitude;  these particles go through our atmosphere reaching deep layers below the ground. For our analysis regarding the Phosphorus characterization, we take into account the most relativistic particles at sea level, that is muons and muon produced photons. Viruses are obligate intracellular parasites composed of a nucleic acid surrounded by a protein coat, the capsid. Some viruses contain a lipid envelope, derived from the host, surrounding the capsid. The nucleic acid found in viruses can consist of either RNA or DNA. The Coronaviridae, for example contain a single molecule of RNA consisting of about 30Kb [21]. RNA is composed of nucleotides, each containing a sugar (deoxyribose), a Nitrogen containing Base (Adenine, Uracil, Guanine, and Cytosine), and a phosphate group $PO_4$. Members of the family Coronoviridae measure 80-160 nm in diameter. Ambient bacteria measure more than one μm in diameter, and containing different chemical components in fraction term, thus giving us the ability to distinguish between the two weighting the different spectrum and flux contribution. It is here, in the phosphate group where most of the phosphate is located. There will also exist trace amounts of phosphate in viral proteins that contain the amino acid methionine. Hence, the vast majority of Phosphorus can be found in the genetic strand of RNA.

Consequently as a benchmark test we assumed as a marker of no living matter ( potentially a virus ) the Phosphorus present in the RNA, which, must be emphasized, is absent in the EPA pollutants list Table 1, and we use only Phosphorus just as a stress test to evaluate if our physical model and particle code were able to detect an isotope with concentrations from 1 ppm in air ( that means one atom of Phosphorus per one million of air molecules ) to 1.%, according to a composition of air and a simplified chemical form of a virus as a biological matter in air as reported in Fig 1 [1,2,3,4,5,6,16,21] with its own atom modelling.

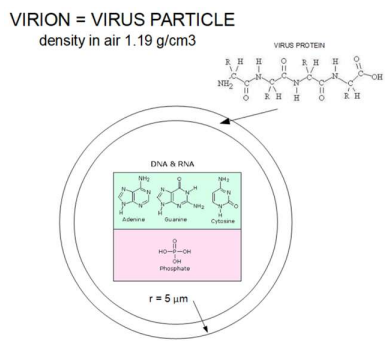

**Figure 1 Particle**

PM and VOC*
NOX
SOX
CO
O3

*PM and VOC

| | | | |
|---|---|---|---|
| 1,1,1-Trichloroethane | Bromomethane | 1,3-Butadiene | Formaldehyde |
| 1,1,2,2-Tetrachloroethane | Carbon Disulfide | 2-Butanone | Gasoline, Automotive |
| 1,1,2-Trichloroethane | Carbon Tetrachloride | 2-Hexanone | Hexachlorobutadiene |
| 1,1-Dichloroethane | Chlorobenzene | Acetone | Hexachloroethane |
| 1,1-Dichloroethene | Chloroethane | Acrolein | Hydrazines |
| 1,2,3 Trichloropropane | Chloroform | Benzene | Methyl Mercaptan |
| 1,2-Dibromo-3-Chloropropane | Chloromethane | Bromodichloromethane | n-Hexane |
| 1,2-Dibromoethane | Dichlorobenzenes | Stoddard Solvent | Nitrobenzene |
| 1,2-Dichloroethane | Dichloropropenes | Toluene | Styrene |
| 1,2-Dichloroethene | Ethylbenzene | Trichloroethylene (TCE) | Tetrachloroethylene (PERC) |
| 1,2-Dichloropropane | Ethylene Dibromide | Vinyl Chloride | Xylenes |

**Table 1 Air Model**

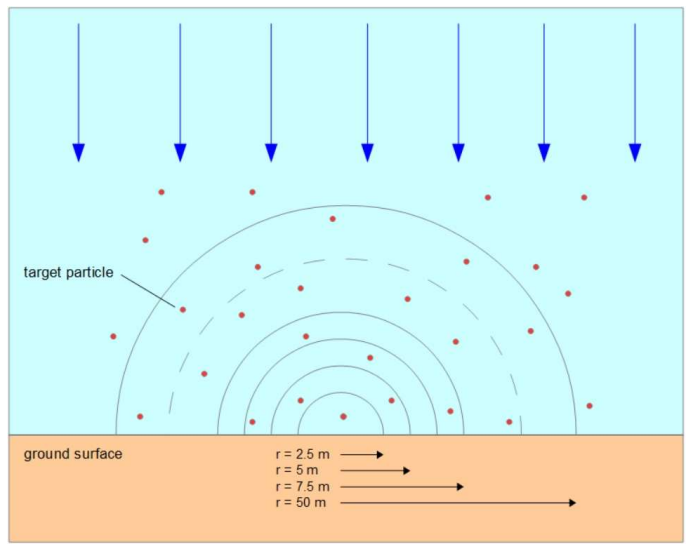

**Figure 2 Physic Model**

## 2. Assumptions & Calculations.

A simulation has been performed by means of the Monte Carlo particle code MCNPX [7] with the particles experimental cross section library ENDF/B-VI [7,8,9].

The particles source has been evaluated at different elevations as a point source shooting muon to the ground starting from a height of 50 m down to 2.5 m as shown in Figs 2-3-4 [10]. For the purpose of this study we conducted all our analyses at 2.5 m where the potential virus particles could affect a human being by inhalation.

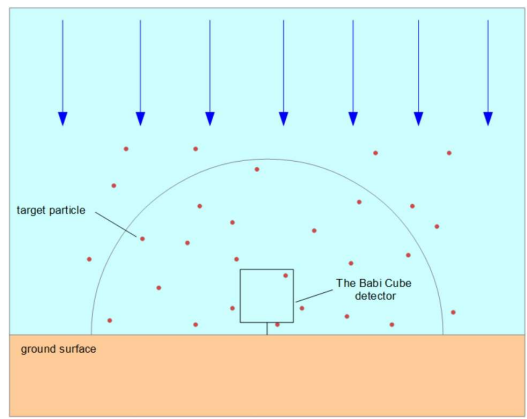

**Figure 3 Detection Model**

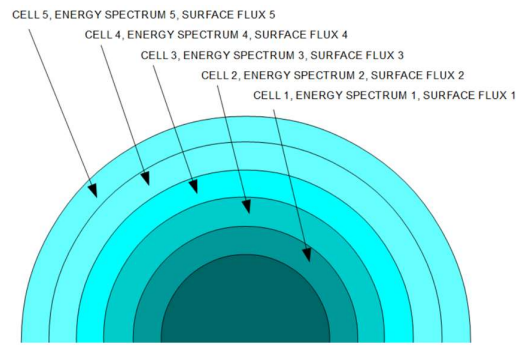

**Figure 4 Geometry Model**

The source of muons at sea level is commonly known to be, approximately and on average, one muon per square centimeter per minute. The muons energy spectrum is shown in fig. 5 (ref. 11,12,19,20).

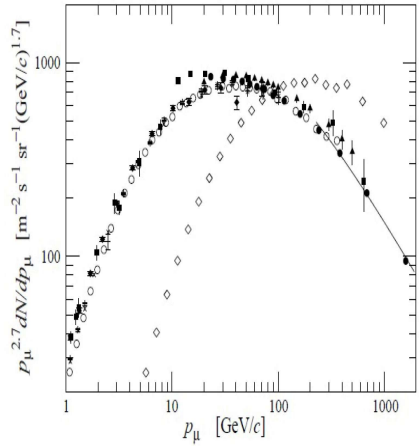

**Figure 5 Muon spectrum (left zenith angle = 0°, right zenith angle= 75°)**

The composition of the air assumed in our model is taken from the official EPA and CDC released data [13,14,15], with all the possible pollutants and contaminants taken into consideration. The composition of air contaminated by an amount of no living matter (actually a virus) is reported in Table 2 [1,2,3,4,5,6,16]. Different concentrations of chemical elements in no living matter (hereafter identified as Phosphorus) have been taken into account, from 1 ppm to 1. % . As a benchmark test we decided to take in consideration a cinematic scenario where we have considered a potential virus load source [17,18] coming from a fixed and constant 100 person infected enclosed in a public open environment and testing different type of virion concentrations to evaluate the detection feasibility process. We took in consideration for it a hall of 6,250 m3 and of dimensions 50m(L) X 50m(W) X 2.5m(H) with a loading factor of 10 hours.

Muons and only muon produced photons, between the vast number of other particles, have been simulated.

Muons and photons fluxes (particles per square centimeter per second) at different elevations have been calculated and the results, as a function of different Phosphorus contamination percentages, are shown in Figs 6-7-8-9-10-11-12-13-14-15 (see at chapter 3 the discussion of the results). It has to be pointed out that multiple techniques have been used in MCNPX to get a good statistic; actually, the error associated to the numbers of the above figures are of the order of 1% or less. It also has to be underlined that even if the conditions of the data taken into account are at sea level, the results are still significant for higher elevations since the density of the air decreases, so that the interactions of cosmic rays decrease, but the amount of available cosmic rays increases so that to counterbalance the first effect.

|  |  | Atomic Fr | % |
|---|---|---|---|
| **Virus Protein Surface Model** | H | 10 | 0.37037037 |
|  | O | 5 | 0.185185185 |
|  | C | 8 | 0.296296296 |
|  | N | 4 | 0.148148148 |
| Norm |  | 27 | 100% |
|  |  |  | % |
| **Core Particle Model** | H |  | 0.494590323 |
|  | O |  | 0.198958065 |
|  | C |  | 0.080645161 |
|  | P |  | 0.016129032 |
|  | N |  | 0.209677419 |
| Norm |  |  | 100% |

**Table 2 Virus in Air Composition**

## 3. Results & Discussion

In this chapter, we discuss the results of our analysis showing the fluxes and energy spectra of our simulation Figs 6-7-8-9-10-11-12-13-14-15).

The graphs below are showing the energy spectra and their different behaviours at 73 discrete energy bins for muon and photon particles.

The graphs Figs 6-7-8-9-10-11-13 have not been normalized, whereas the graphs Figs 12-14-15 have been normalized; particle fluences have been estimated on a sample detector design ( "The Babi Cube" which has an acquisition volume of $1m^3$ and the physical phenomena have been evaluated at different exposure times .

As far as the photon behaviour is involved, an interesting energy spectrum amplifier phenomenon located at 0.05, 0.2, 2., 100. MeV can be seen and considered as a unique signature in contamination percentages, especially as reported in Fig-6-8-9-10-11.

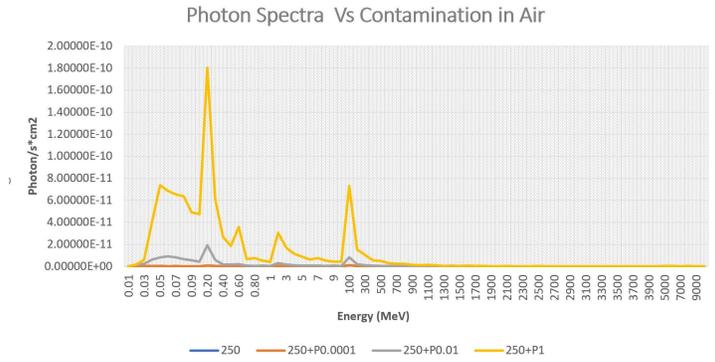

**Figure 6 Photon fluxes as a function of energy for different Phosphorus percentages**

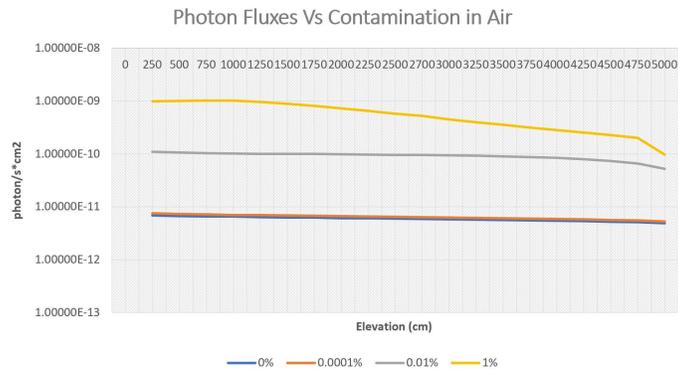

**Figure 7 Photon fluxes as a function of elevation for different Phosphorus percentages**

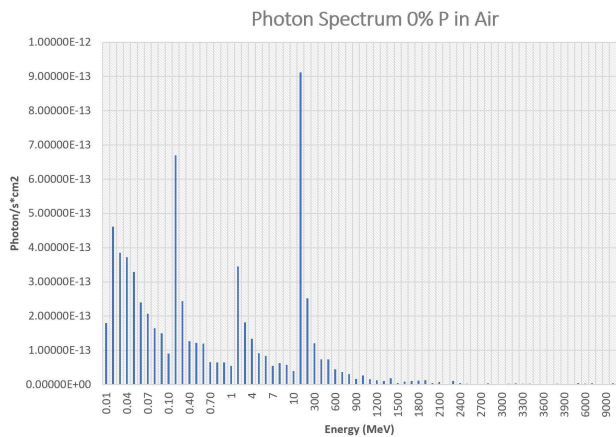

**Figure 8 Photon fluxes as a function of energy with no Phosphorus contamination**

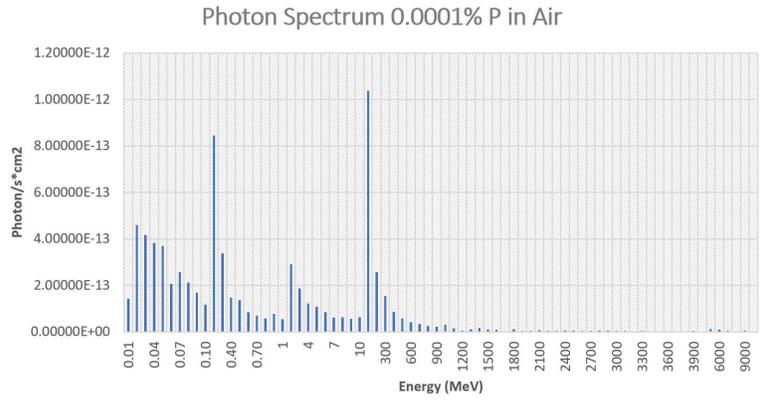

**Figure 9 Photon fluxes as a function of energy with 1. ppm Phosphorus**

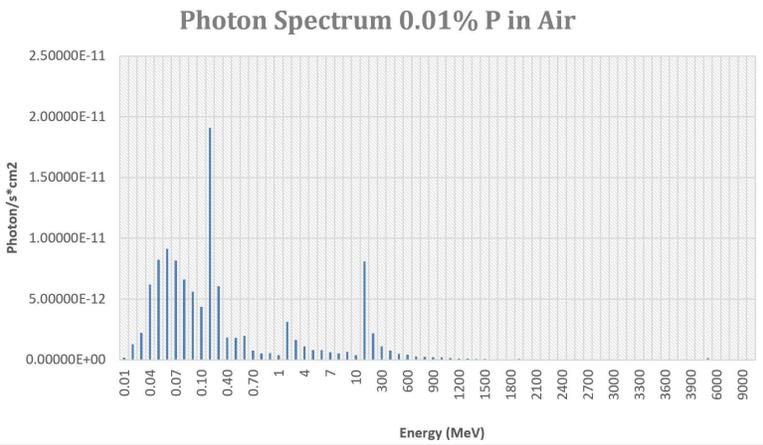

**Figure 10 Photon fluxes as a function of energy with 100. ppm Phosphorus**

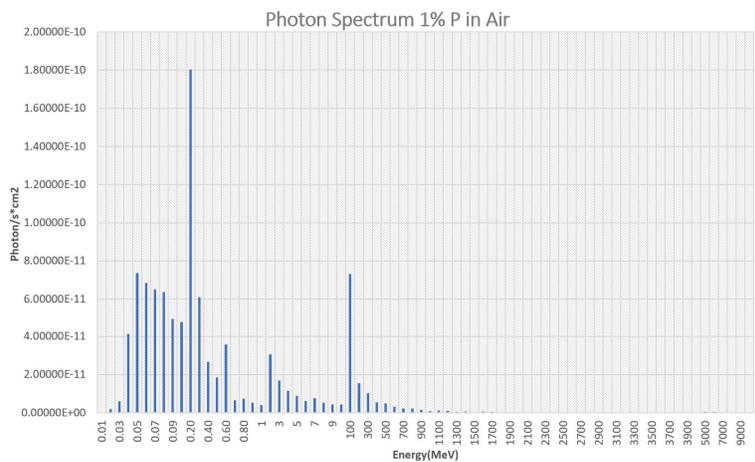

**Figure 11 Photon fluxes as a function of energy with 1% Phosphorus**

In open air, the photon contribution would need an imposed photon source in order to scaling up to detector fluence Fig 12. However, in closed areas its contribution could be effective due to the interaction with higher atomic elements present in the surrounding buildings' perimetry

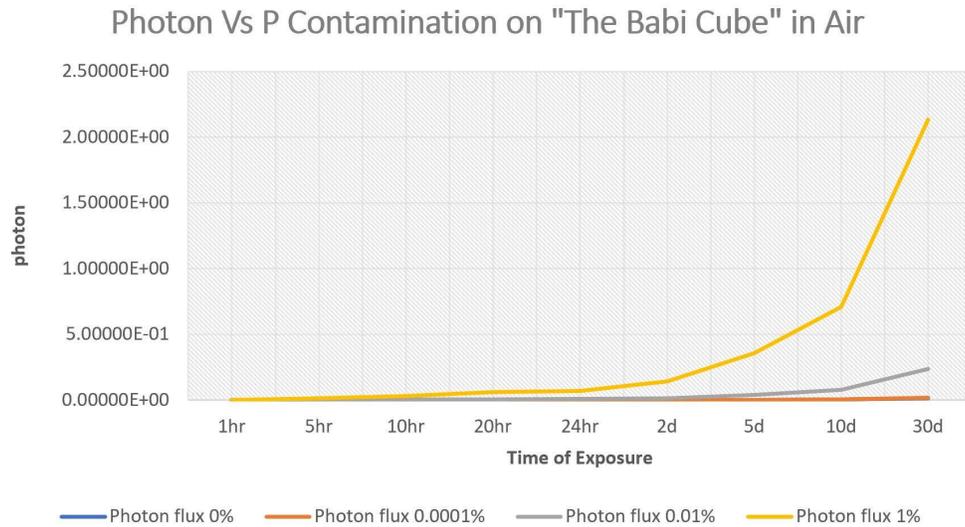

Figure 12 Photon fluences as a function of time for different Phosphorus percentages

Per contra the muon behaviour shows an evolution which can be significant to discriminate the concentration of Phosphorus in air.

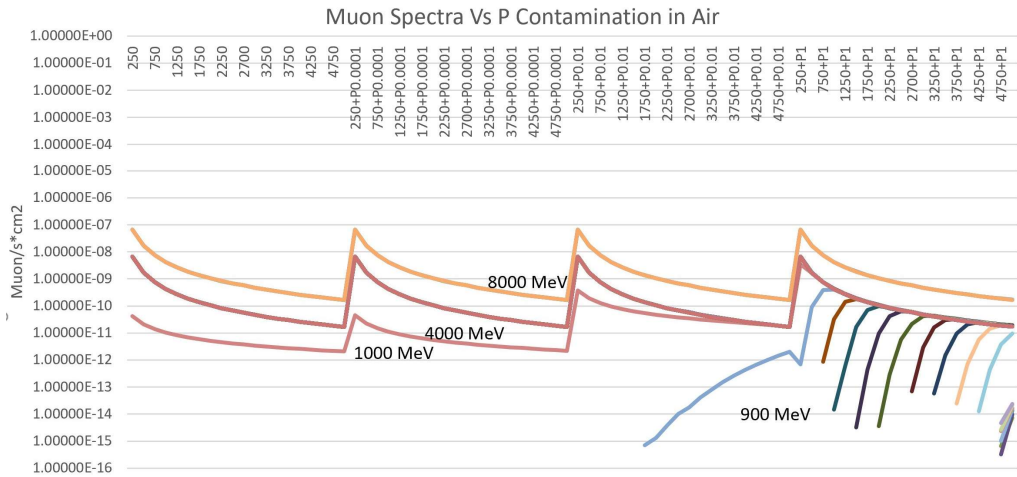

Figure 13 Muon fluxes as a function of elevation for Phosphorus percentages at different energies

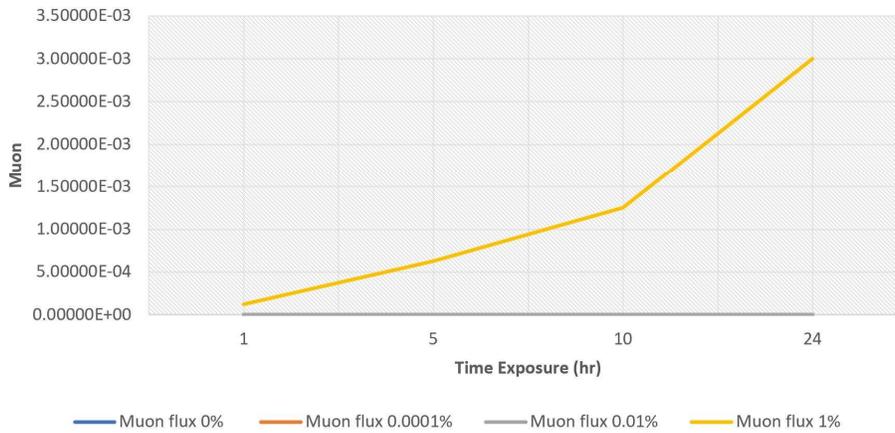

Figure 14 Muon fluences as a function of time for different Phosphorus percentages at 0.9 GeV

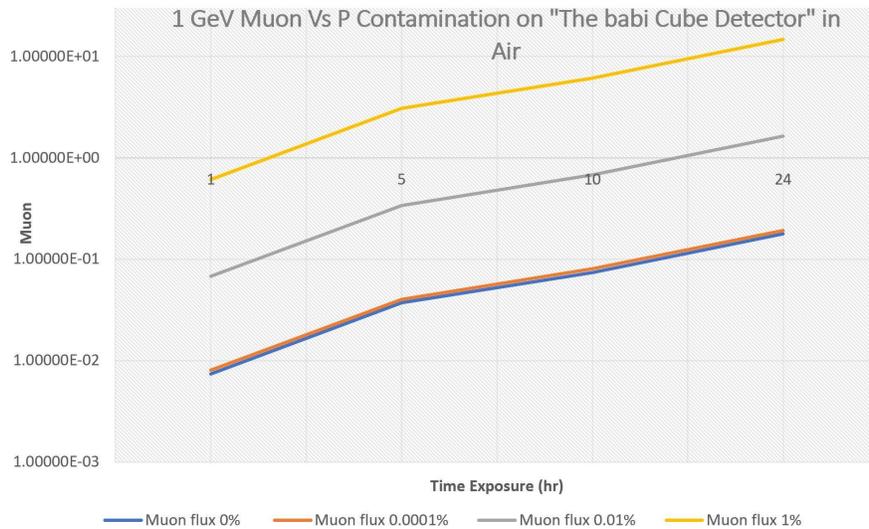

**Figure 15 Muon fluences as a function of time for different Phosphorus percentages at 1. GeV**

Its spectrum (Fig 13) shows an interesting behaviour located at 0.9, 1.0 and the <900 MeV, values within the minimum and maximum 100 MeV – 8 GeV. All those spectrum energy lines are acting as unique identifier for the different contamination present in the air allowing to discriminate the muons as a function of the Phosphorus concentrations in open air. In Fig 13 the 0.9 GeV channel is a unique contamination marker for 0.01% and 1% cases. Furthermore, it has to be highlighted that the channel 0.9 GeV is remarkable in the 1% case in term of fluence (Fig 14).

The 0.9-1 GeV and low band energy cascade channels are markers for 0%, 0.0001%, 0.01% and 1% showing different flux peaks. (Figs 15)

Only the case 1% shows a low energy spectrum cascade starting from 800 MeV to 100 MeV. (Fig 13)

## 4. Summary

The study research proposed shows a potential new methodology approach regarding a physical model and its simulation performed by the Los Alamos National Laboratory MCNPX particle code demonstrate it's possible to identify low levels of contamination of Phosphorus (present in no living matter nucleic acid) mixed in air thus aiding in evaluating the energy spectra and associated particle fluxes of cosmic muons and muon produced photons.

It's really interesting the photon spectra and fluxes however that would require a impose external source for detection.

Regarding in particular the muon energy spectrum is potentially able to discriminate the different contamination concentrations thanks to 0.9 GeV, 1 GeV peaks and the low band <900 MeV present in the muon flux showing a unique trend pattern as a function of the contamination percentage in air. Every single contamination is unique in its own "spectrum particle signature" acting as a unique identifier in the detection process. Moreover, the discrimination ratio fm/fp (where fm and fp stand for muon and photon fluxes) can also be an "add on" technique to identify the contamination case under investigation and detectors can accordingly alarm its presence in the ambient with the potential ability to discriminate what kind of virus is present in the air.

It's also has to be underlined that the particle source (cosmic rays) cannot be altered or modified in order to falsify the ppm outcome of viruses in air, in other words, is a universal not disputable (and free) reference.

## 5. Outlook

Our team is designing an appropriate detector able to discriminate the particle spectra and fluxes. Its most relevant characteristics are an extremely large sensitive area and a high resolution for the discrimination of muons of energies significant for the determination of the presence of viral phosphorus from other radiation components. A specific study for the definition of the detector parameters is in progress.

Moreover, our current research activity is focusing, with our Virology team, on a more complex virus characterization in order to discriminate different types of viruses such as Covid19, Ebola, common Flu.

Our Initial research approach and intention were studying and evaluating the feasibility of cosmic rays interactions in order to detect viral particles as we reported in our results. For that purpose and intention, the mixture homogenization process was acceptable to evaluate the $PO_4$ group present in different quantities (Kb strands) as a function of the virus type under study.

For the discrimination porpoise we will describe every virus in a cluster configuration through a volumetric matrix grid cell in function of: type of encoded protein surface, type of core RNA-DNA strands and different Kb, leading to have a non-homogeneous medium giving us the ability to analyze, identify and discriminate our studying object through a spectrum analysis.

Moreover, because the secondary particle flux composed by photons(generated by muons) was too low in term of fluence we will studying also an external imposed photon source interacting with the same matrix cluster configuration in order to detect and discriminate the type of virus indoor too or outdoor.

The main reason about the choice of cosmic rays as a detection source is based not only on the fact that is a "free and undisputable reference" but also on the public safety/shielding requirements due to external X-rays bursts coming from an artificial potential photon source imposed.

**Acknowledgements**

We deeply thank, Dr. Ilaria A. Valli.

# References

1. D.G Sharp, A.R Taylor, I.W McLean Jr, Dorothy Beard, J.W Beard, *Density and Size of the Influenza Virus A (PR8 Strain) and B (Lee Strain) and the Swine Influenza Virus,* 1945

2. Kerby F. Fannin, Stanley C. Vana, Walter Jacubowski, *Effect of an Activated Sludge Wastewater Treatment Plant on Ambient Air Densities of Aerosols containing Bacteria and Viruses,* 1985

3. Wan Yang, Subbiah Elankumaran, Linsey C. Marr, *Concentrations and size distributions of airborne influenza A viruses measured indoors at a health centre, a day-care centre and on aeroplanes,* 2011

4. Nicole M. Bouvier, Peter Palese, *The Biology of Influenza Viruses,* 2008

5. Jeffrey E Lee, Erica Ollmann Saphire, *Ebolavirus glycoprotein structure and mechanism of entry,* 2010

6. Xurong Wang, Fuxian Zhang, Rui Su, Xiaowu Li, Wenyuan Chen, Qingxiu Chen, Tao Yang, Jiawei Wang, Hongrong Liu, Qin Fang, and Lingpeng Cheng, *Structure of RNA polymerase complex and genome within a dsRNA virus provides insights into the mechanisms of transcription and assembly,* 2018

7. Denis B. Pelowitz, Los Alamos National laboratory, *MCNPX*, 2005 LA-CP-05-0369

8. Morgan C. White, Los Alamos National Laboratory, *Photo atomic Data Library MCPLIB04, 2003*

9. Oak Ridge National Laboratory, *MCNP-MCNPX Code Collection, 2006*

10. John W. Watts Jr, NASA, *Calculation of Energy Deposition for simple Geometries,* 1973

11. Prashant Shukla, Sundaresh Sankrith, *Energy and angular Distribution of Muons at Earth,* 2018

    **12.** Antonio La Calamita, Francesco Lo Parco, *Misure del Rate Giornaliero dei muoni nie raggi cosmici secondari in laboratorio,* 2017

13. United States Environmental Protection Agency, https://www.epa.gov/pmpollution/particulate-matter-pm-basics

14. US Agency For Toxic Substances & Disease Registry https://www.atsdr.cdc.gov/substances/toxchemicallisting.asp?sysid=7

15. US Center For Disease Control and Prevention, https://www.cdc.gov/air/particulate_matter.html

16. Owen Pornillos, Barbie K. Ganser-Pornillos & Mark Yeager, *Atomic-level modelling of the HIV capsid,* 2011

17. Nikolai Nikitin, Ekaterina Petrova, Ekaterina Trifonova, and Olga Karpova, *Influenza Virus Aerosols in the Air and Their Infectiousness,* 2014

18. J.R. Brown a, J.W. Tang b, *, L. Pankhurst c, N. Klein d, V. Gant e, K.M. Lai f, J. McCauley g, J. Breuer, *Influenza Virus Survival in Aerosols and estimates of viable virus loss resulting from aerosolization and air-sampling, 2015*

19. C. L. Morris, Konstantin Borozdin, Jeffrey Bacon, Elliott Chen, Zarija Lukić, Edward Milner, Haruo Miyadera, John Perry, Dave Schwellenbach, Derek Aberle, Wendi Dreesen, J. Andrew Green, George G. McDuff, Kanetada Nagamine, Michael Sossong, Candace Spore, and Nathan Toleman, *Obtaining material identification with cosmic ray radiography*, 2012

20. G. Bonomi, P. Checchia, M. D'Errico, D. Pagano, G. Saracino*, Application for Cosmic Rays*, 2020

21. Sah, Rabjit et. al., Microbiology Resource Announcement, American Society for Microbiology, 2020